\documentclass[final]{svjour2}
\usepackage{graphicx}
\usepackage{rotating}
\usepackage{amssymb}
\usepackage{mathptmx}
\usepackage[square,numbers,sort&compress,merge]{natbib}
\usepackage{epstopdf}
\usepackage{textcomp}
\usepackage{fixltx2e}
\usepackage{hyperref}
\makeatletter
\journalname{Journal of Low Temperature Physics}

\begin{document}

\newcommand{\hdblarrow}{H\makebox[0.9ex][l]{$\downdownarrows$}-}
\title{Optical Demonstration of THz, Dual-Polarization Sensitive Microwave Kinetic Inductance Detectors}

\author{B.~Dober\textsuperscript{1} \and J.A.~Austermann\textsuperscript{2} \and J.A.~Beall\textsuperscript{2} \and D.~Becker\textsuperscript{2} \and G.~Che\textsuperscript{3} \and H.M.~Cho\textsuperscript{4} \and M.~Devlin\textsuperscript{1} \and S.M.~Duff\textsuperscript{2} \and N.~Galitzki\textsuperscript{1} \and J.~Gao\textsuperscript{2} \and C.~Groppi\textsuperscript{3} \and G.C.~Hilton\textsuperscript{2} \and J.~Hubmayr\textsuperscript{2} \and K.D.~Irwin\textsuperscript{4} \and C.M.~McKenney\textsuperscript{2} \and D.~Li\textsuperscript{4} \and N.~Lourie\textsuperscript{2} \and P.~Mauskopf\textsuperscript{3} \and M.R.~Vissers\textsuperscript{2} \and Y.~Wang\textsuperscript{2}}

\institute{1: University of Pennsylvania, Philadelphia, PA \\ 2: National Institute of Standards and Technology, Boulder, CO \\ 3: Arizona State University, Tempe, AZ \\ 4: Stanford University, Stanford, CA \\ 
\email{dober@sas.upenn.edu}}

\maketitle

\begin{abstract}

Polarization-sensitive, microwave kinetic inductance detectors (MKIDs) are under development for the next-generation BLAST instrument (BLAST-TNG). BLAST-TNG is a balloon-borne submillimeter polarimeter designed to study magnetic fields in diffuse dust regions and molecular clouds. We present the design and performance of feedhorn-coupled, dual-polarization sensitive MKIDs fabricated from TiN/Ti multilayer films, which have been optimized for the 250 $\mu$m band. Measurements show effective selection of linear polarization and good electrical isolation between the orthogonally crossed X and Y detectors within a single spatial pixel. The detector cross-polar coupling is $<$3\%. Passband measurements are presented, which demonstrate that the desired band-edges (1.0-1.4~THz) have been achieved. We find a near linear response to the optical load from a blackbody source, which has been observed in previous devices fabricated from TiN. Blackbody-coupled noise measurements demonstrate that the sensitivity of the detectors is limited by photon-noise when the optical load is greater than 1~pW.



\end{abstract}

\section{Introduction}
 The next generation BLAST experiment~\citep{dober, tyr} (BLAST-TNG) is a suborbital balloon payload that seeks to map polarized dust emission in the 250~$\mu$m, 350~$\mu$m and 500~$\mu$m wavebands. The instrument utilizes a stepped half-wave plate to reduce systematics. The general requirement of the detectors is that they are photon-noise-limited and dual-polarization sensitive. To achieve this goal, we are developing three monolithic arrays of cryogenic sensors, one for each waveband. Each array is feedhorn-coupled and each spatial pixel consists of two orthogonally spaced polarization-sensitive microwave kinetic inductance detectors~\citep{Day2003} (MKIDs) fabricated from a Ti/TiN multilayer film. In previous work, we demonstrated photon-noise-limited sensitivity in 250~$\mu$m waveband single polarization devices~\citep{hubmayr2015}. In this work, we present the first results of dual-polarization sensitive MKIDs at 250~$\mu$m.

\section{Detector Design}
\label{sec:det}
Our detection scheme utilizes feedhorn/waveguide, front-side optical coupling and places the inductive section of a lumped-element kinetic inductance detector~\citep{Doyle} one-quarter wavelength away from a reflective backshort.  This coupling approach is described in detail in Hubmayr et al~\citep{hubmayr2015}.  

The devices are fabricated from TiN/Ti proximitized films, which have a tunable and spatially uniform T{\scriptsize c}~\citep{Vissers}. The TiN(Ti) thickness is 4(10) nm, which sets T{\scriptsize c}=1.35~K. By stacking a number of bilayers, we may tune the sheet resistance of the film without altering T{\scriptsize c}. We chose to stack four bilayers and add a protective TiN cap layer, which produces $R_s$~= ~20 $\Omega / \Box$ and $L_s$~=22.5 pH/$\Box$. The TiN/Ti multilayer effectively reduces the sheet impedance by a factor of four as compared to the TiN/Ti/TiN trilayer films used in our previous single-pol devices~\citep{hubmayr2015}. This allows us to reduce the absorber width by a factor of four in the new dual-polarization devices, which is critical to minimize their cross-polar coupling.

\begin{figure}
\centering
\includegraphics[width=1.0\linewidth]{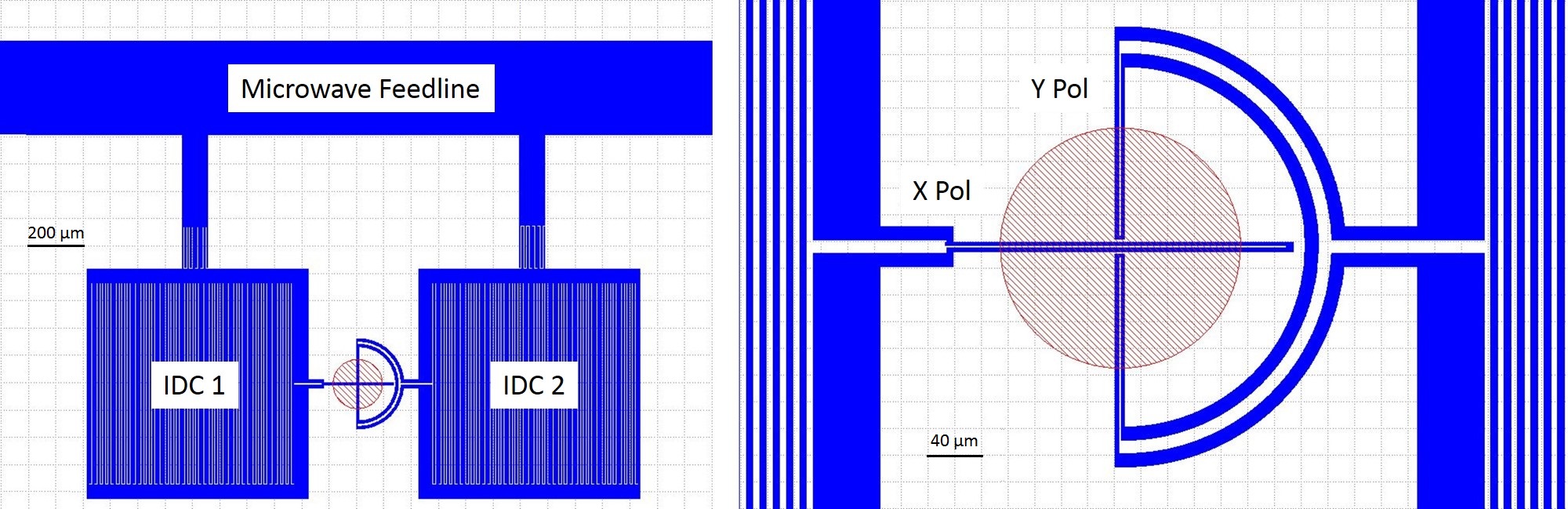}
\caption{Detector Design. \textit{Left:} An overview of a single pixel. The microstrip feedline at the top of the figure is capacitively coupled to the two X and Y polarization lumped-element MKIDs. The large interdigitated capacitor comprises the majority of the MKID. The 180~$\mu$m diameter waveguide which illuminates the inductors is depicted by the shadowed circular region. \textit{Right:} A magnification of the inductive meanders which act as the polarization-sensitive absorbers. The two 3.2~$\mu$m thick inductors are non-intersecting with the Y polarization detector ending 2~$\mu$m before intersecting the X polarization detector. (Color figure online.)}
\label{fig:NonCrossOverLayout}
\end{figure}

Fig.~\ref{fig:NonCrossOverLayout} shows the photolithography mask design used to produce \\ dual-polarization-sensitive MKIDs from these films. There are two MKIDs per spatial pixel, one per linear polarization. Each MKID contains a 5~$\mu$m finger/gap interdigitated capacitor of total area 0.68~mm$^2$ and a 3.2~$\mu$m wide inductor that spans the length of the 180~$\mu$m wide waveguide diameter for a total volume of 154~$\mu$m$^3$ and 230~$\mu$m$^3$ for the X and Y inductors, respectively. This combination of $L$ and $C$ produces resonance frequencies near $f_o$~$\sim$~1~GHz.  Each MKID couples to a 340~$\mu$m wide microstrip transmission line (the silicon wafer is the dielectric and the device box is the ground plane) via an interdigitated coupling finger of designed $Q_c$ $\sim$ 40000-50000.     

The two inductors within a pixel are orthogonally aligned in order to obtain dual-polarization-sensitivity. By making the Y-polarization inductor discontinuous, both MKIDs are defined in one device layer without requiring electrical cross-overs. In electro-magnetic simulations of a simplified model of just the antennas, the inherent asymmetry of the design produces a band averaged (1-1.4~THz) co-polar coupling of 79 (75)\% in the continuous (discontinuous) absorbing inductor. These simulations also suggest the expected cross-polar coupling to be $<$2\%, but if the vacuum gap between the wafer and the feedhorn/waveguide becomes too large, other structures in the detector design could begin to produce an additional cross-pol contribution.

Five-pixel prototype arrays have been fabricated on a $2f\lambda$ (2.5~mm) detector pitch.  An array couples to a matching array of aluminum, direct-machined feedhorns. The horns are a three step modified Potter horn that has been designed for minimized beam asymmetries while achieving a 30\% fractional bandwidth~\citep{Potter}\citep{Zeng}. The 1.0~THz low edge of the band is defined by waveguide. A quasi-optical low-pass filter mounts in front of the feedhorns, which defines the 1.4~THz high edge of the passband~\citep{Ade}. We mount this detector package to the cold stage of an adiabatic demagnetization refrigerator and operate the array at 100~mK in the measurements described below.   

\section{Polarization Response and Passbands}
\begin{figure}
\centering
\includegraphics[width=0.7\linewidth]{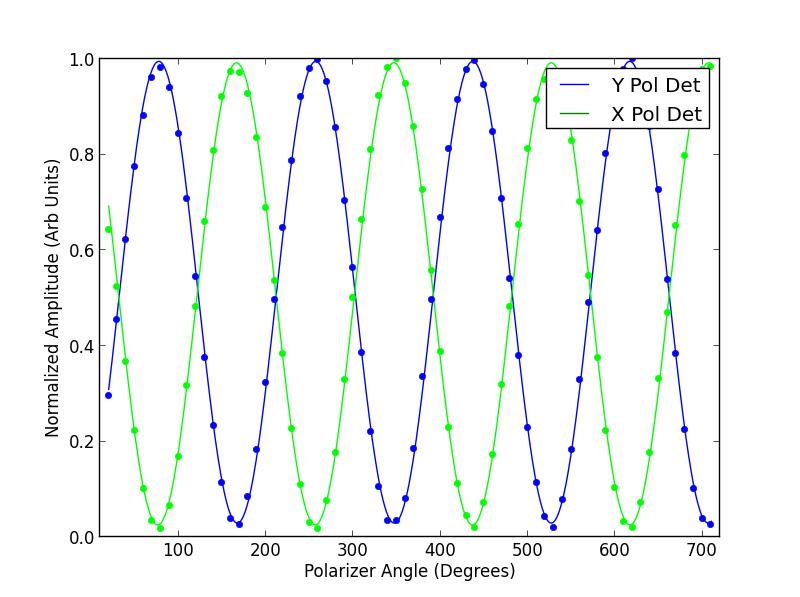}
\caption{Polarization Efficiency Graph. The data points are the amplitudes of the chopped signal at each input polarization angle, normalized to the peak signal, while the lines are a fit to the data. The resulting cross-polar coupling is at most $2.6\%$ and $2.8\%$ for the X and Y polarization detectors, respectively. (Color figure online.)}
\label{fig:poleff}
\end{figure}

Polarization characterization was performed on the prototype array described in Section~\ref*{sec:det} in a cryostat that is optically coupled to the room with appropriate quasi-optical filtering. In addition, a 1.8~mm thick piece of eccosorb MF-110 microwave absorber was installed to decrease the optical loading on the detectors, ensuring their operability when viewing a 300~K thermal load. The microwave absorber has an anti-reflective coating and has a calculated band-averaged transmission of 0.93\%. The detectors are coupled to a 1050~C to 20~C chopped thermal source that underfills the beam of the feedhorns. A rotatable wire grid polarizer which has an induced cross-pol of less than 0.5\% is placed between the chopped source and the cryostat window. We determine the polarization properties by measuring the amplitude of the response of the detectors as a function of the angular position of the polarizer. The result produces a sinusoidal signal, shown in Fig.~\ref{fig:poleff}. We determine the cross-polar coupling, or the minimum of the amplitude response, by fitting the data to a sine wave. The results of the fit suggests the detector cross-polar coupling is 2.6\% and 2.8\% for the X and Y polarizations, respectively. However, these values are most likely an upper limit as the fit lies above the minimum data points (1.7\% and 1.9\% for X and Y, respectively). Regardless, this result is consistent within the uncertainty of the HFSS simulations for this pixel geometry, and is an improvement over the 11\% cross-polar coupling measured in the previous experiment\cite{enzo}. 

\begin{figure}
\centering
\includegraphics[width=0.7\linewidth]{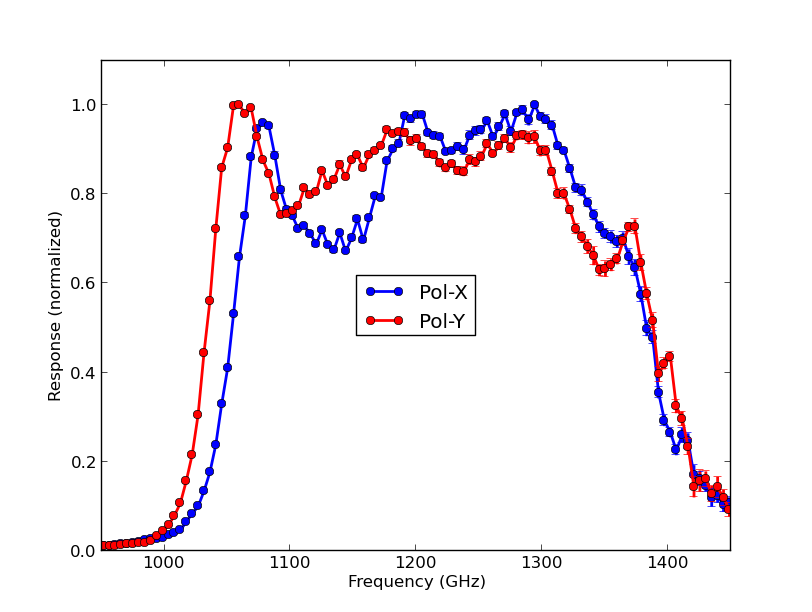}
\caption{Detector Bandpasses. The X and Y polarization bandpasses, averaged over multiple FTS measurements, are corrected for the spectral slope in the transmission of the microwave absorbing filter. The uniform high frequency cutoff is defined by a low-pass filter mounted directly on top of the feedhorns. The low frequency cut-on is defined by the feedhorn waveguide diameter. The non-uniformity in X and Y polarization detectors is due to the slight ellipticity in the waveguide due to machining techniques and will be addressed in the final feedhorn array. (Color figure online.)}
\label{fig:bandpass}
\end{figure}

We determine the detector passbands by use of a Fourier Transform Spectrometer (FTS), which was purged with nitrogen gas to minimize atmospheric attenuation. The FTS is coupled to the same 1050~C thermal source used in the polarization measurements. In this configuration, the source fills the beam and is no longer chopped, and the input polarizer is removed. The bandpasses for each X and Y polarization detector are averaged over multiple FTS scans to reduce spectral noise, and the result is shown in Fig.~\ref{fig:bandpass}. While the cut-off frequency (1.4~THz), which is defined by the low-pass filter, is uniform, there is a clear difference in cut-on frequencies between the two detector polarizations. The X polarization cut-on is 1054.1~GHz, while the Y polarization cut-on is 1033.7~GHz. This 20.4 GHz discrepancy in the low frequency edge is likely due to a slightly oval-shaped waveguide (3.55 microns larger in Y-pol than X-pol), which is produced by using a standard twist drill bit. To address this problem, the waveguide will be undercut and reamed to produce a uniform circle at the desired waveguide diameter.  

\section{Detector Thermal Blackbody Responsivity}
\begin{figure}
\centering
\includegraphics[width=1.0\linewidth]{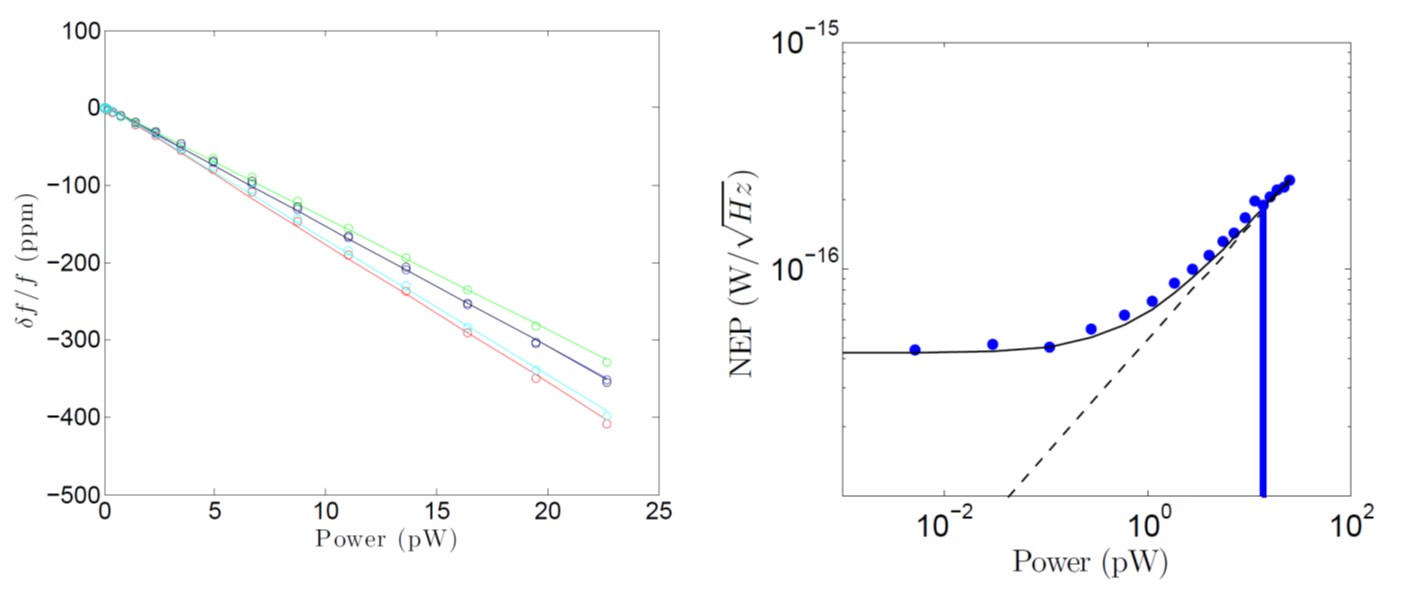}
\caption{Detector Responsivity and Sensitivity. The horizontal axis of both plots is power emitted from the blackbody load, which is calculated based on the temperature of the blackbody load using the Planck function and assuming single-mode, single-polarization coupling from 1.0-1.4 THz. \textit{Left:} A sample of responsivities from the 5 pixel prototype array. These devices show a near linear responsivity to photon load. \textit{Right:} The noise equivalent power (NEP) of the detectors at a range of blackbody temperatures from 3~K-22~K. The blue points are noise data taken at increasing blackbody temperatures, while the dashed line is the blackbody's expected NEP. The black curve is a fit to the NEP of the detector which is used to calculate and optical coupling efficiency of $\sim$75\%. The expected photon noise of the instrument for the 250$\mu$m array is shown as the blue bar at 13.8~pW. (Color figure online.)}
\label{fig:nep_dualpol}
\end{figure}

We determine the sensitivity to thermal radiation by coupling the detectors to a beam-filling temperature-controlled blackbody load. The measurement approach is identical to that described in Hubmayr et al~\citep{hubmayr2015}. While the detectors were held at a bath temperature of 100~mK, the frequency noise and detector response was measured as a function of blackbody temperature 3~K to 22~K. Our previous trilayer films~\citep{hubmayr2015} as well as other devices fabricated from TiN~\citep{McKenney2012, Omid} show a near linear responsivity to photon load. We observe the same phenomenon in these multilayer films as seen in Fig.~\ref{fig:nep_dualpol}. The device responsivity ranges from -20 to -24 ppm/pW for a sample of MKIDs in the 5-pixel array. The frequency noise of the detector taken at each temperature step is converted to a noise equivalent power (NEP) by utilizing the local slope of the detector responsivities in Fig.~\ref{fig:nep_dualpol}. The results are fitted using a best fit NEP model which accounts for individual sources of noise. More detail on this can be found in Hubmayr et al~\citep{hubmayr2015}. This best fit model is used to determine a detector optical coupling efficiency of $\sim$75\%. These results also confirm that the multilayer films demonstrate photon noise limited sensitivity above 1~pW of loading.

\section{Conclusion}
We demonstrate dual-polarization sensitive THz pixels comprised of microwave kinetic inductance detectors. These detectors have good polarization isolation, demonstrating a cross-polar coupling of at most 2.6\% and 2.8\% in the X and Y polarizations, respectively. In addition, these detectors demonstrate background limited sensitivity above 1~pW of loading. This work represents a viable path towards production of large arrays of thousands of polarization sensitive MKIDs for observations in the sub-mm which will be deployed on the BLAST-TNG instrument. 

\begin{acknowledgements}
This work was supported in part by NASA through Grant No. NNX13AE50G. Brad Dober is supported through the NASA Earth and Space Science Fellowship (Grant No. NNX12AL58H).
\end{acknowledgements}

\end{document}